\documentclass{aastex6}
\usepackage{mathrsfs}
\usepackage{amsmath}
\usepackage{color}
\usepackage{xcolor}
\begin{document}

\title{Numerical Simulations of a Jet-Cloud Collision and Starburst:  Application to Minkowski's Object}
\shorttitle{Simulations of a Jet-Cloud Collision}
\shortauthors{Fragile, et al.}
\author{P. Chris Fragile}
\affil{Department of Physics \& Astronomy, College of Charleston, Charleston, SC 29424, USA}
\email{fragilep@cofc.edu}
\author{Peter Anninos}
\affil{Lawrence Livermore National Laboratory, Livermore, CA 94550, USA}
\author{Steve Croft}
\affil{Astronomy Department, University of California, Berkeley, 501 Campbell Hall \#3411, Berkeley, CA 94720, USA}
\affil{Eureka Scientific, Inc. 2452 Delmer Street Suite 100, Oakland, CA 94602, USA}
\author{Mark Lacy}
\affil{National Radio Astronomy Observatory, 520 Edgemont Road, Charlottesville, VA 22903, USA}
\author{Jason W. L.  Witry}
\affil{Department of Physics \& Astronomy, College of Charleston, Charleston, SC 29424, USA}

\begin{abstract}
We present results of three-dimensional, multi-physics simulations of an AGN jet colliding with an intergalactic cloud.  The purpose of these simulations is to assess the degree of ``positive feedback,'' i.e. jet-induced star formation, that results.  We have specifically tailored our simulation parameters to facilitate comparison with recent observations of Minkowski's Object (M.O.), a stellar nursery located at the termination point of a radio jet coming from galaxy NGC 541.  As shown in our simulations, such a collision triggers shocks which propagate around and through the cloud.  These shocks condense the gas and under the right circumstances may trigger cooling instabilities, creating runaway increases in density, to the point that individual clumps can become Jeans unstable.  Our simulations provide information about the expected star formation rate, total mass converted to \ion{H}{1}, H$_2$, and stars, and the relative velocity of the stars and gas.  Our results confirm the possibility of jet-induced star formation, and agree well with the observations of M.O.  
\end{abstract}
\keywords{galaxies: individual (Minkowski's Object) --- galaxies: jets --- hydrodynamics --- intergalactic medium --- shock waves}
\maketitle

\section{Introduction}

The interaction between high-energy jets from active galactic nuclei (AGN) and their surroundings has long been a topic of great astrophysical interest. It is well known that AGN feedback can control the size of a galaxy by influencing star formation, but the mechanism behind this is not well understood.  Several recent observations \citep{Nesvadba10,Guillard15}, as well as numerical studies \citep[e.g.][]{Sutherland07,Antonuccio08,Gaibler12}, have demonstrated that AGN feedback can be either ``negative'' or ``positive.''

The exact astrophysical conditions in the jet and cloud are important in determining what direction feedback takes. Jets can be roughly divided according to their Fanaroff-Riley classification \citep{FR74}. Fast, energetic FRII jets seem more likely to result in negative feedback. Negative feedback curbs or even halts star formation, and is thought to result from the extreme radiative and kinetic energies of the jet, which heat and disperse the star-forming gas. Additionally, the kinetic energy of a jet creates turbulence that can prevent ambient gas from cooling and subsequently coalescing \citep[e.g.][]{Nesvadba10}. For example, a study of the system 3C 326 by \citet{Ogle07} found that, despite the strong H$_{2}$ line emission and an inferred molecular gas mass of 2 $\times 10^{9}$ M$_{\odot}$, the star formation is 20 times lower than predicted by the Kennicutt-Schmidt law.  They infer that turbulent heating from the jet is inhibiting star formation. Other studies, though, suggest increased star formation may be seen in the cocoon of such jets \citep[e.g.][]{Gaibler12}. 

In contrast, FRI jets propagate through the ISM/IGM with energies high enough to create compression in the surrounding gas, but low enough to reduce  the chance of significant turbulent heating. These jets are observed in positive feedback cases,  wherein the effect of the jet serves to enhance star formation, including Centaurus A \citep[][and references therein]{Salome16}, 4C 41.17 \citep{Bicknell00}, and Minkowski's Object (hereafter M.O.) \citep[][hereafter C06]{Croft06}. We give more details on both positive and negative feedback in Section \ref{sec:jetcloud}.  

In this paper, we focus on the case of M.O., a peculiar star forming object located at a redshift of $z = 0.0189$ (C06) that is currently being bombarded by a FR I radio jet from the nearby galaxy NGC 541. The M.O. system is of particular interest due to the lack of evidence for an especially dense ISM or IGM. There is also not much evidence for cold gas outside of the jet interaction site, unlike in Centaurus A. As a result, it is unlikely that significant star formation would proceed in M.O. without the interaction of the jet. 

A strong argument in favor of jet-induced star formation in M.O. is the morphology of the jet-cloud interaction site. Outside of the jet interaction, the gas in M.O. is warm ($\sim 10^4$ K) and clumpy. Near the jet interaction site there is a double structure of \ion{H}{1} gas wrapped around the jet and numerous \ion{H}{2} regions (C06). C06 thus determined that it is likely that the jet interaction in M.O. caused the warm gas to cool into the \ion{H}{1}, in contrast to the pre-existing cold gas regions in Centaurus A. Also, the star forming regions in M.O. correlate with the jet-cloud morphology; the region where the star formation is the highest is the center of the jet-cloud interaction, and the star formation rate (SFR) decreases laterally from this point (C06). 

M.O. may, in fact, be a low redshift example of the type of jet-induced star formation that was perhaps more common in the early universe. Evidence for this is the similarity between M.O. in ultraviolet and the rest-frame UV morphology of suspected jet-induced star-forming regions around high-redshift radio galaxies.

The present work can be viewed as an extension of our earlier study of the interactions of radiative shocks with clouds \citep{Fragile04}. That work focused mostly on the effects of planar shocks overtaking individual (or small collections of) warm clumps on the scale of $\sim 100$ pc.  In the current work, we explore the much richer problem of a full jet intersecting an inhomogeneous intergalactic cloud on the scale of tens of kpc. The paper proceeds as follows: Section 2 covers the theory behind jet-cloud interactions, Section 3 describes the numerical models used to capture the M.O. system, Section 4 details the simulation results, and Section 5 concludes the findings.

\section{Jet-Cloud Interactions}
\label{sec:jetcloud}

The basic idea of jet-induced star formation (i.e. positive feedback) is that the collision of the jet with the cloud will trigger a series of shocks within the cloud.  The immediate effect of these shocks will be to compress and heat the gas. Depending on how the radiative processes scale with density and temperature, the net result can be to dramatically increase the radiative efficiency within the cloud. If the temperature dependence is shallower than the density dependence, the cloud can enter a phase of runaway cooling.  This process occurs most quickly in relatively over-dense regions of the original cloud.  These over-dense regions then proceed to collapse at an accelerating pace.  Provided some of these clumps start sufficiently close to the Jeans limit, this collapse will push them beyond this limit, such that gravitational collapse can take over and the clump will proceed to form stars. 

The properties of the jet and cloud are key to controlling this process.  For positive feedback to be important, the initial cloud must be dense enough for some parts to be reasonably close to the Jeans limit.  The temperature must also be such that any increase in temperature is met with a dramatic increase in cooling (the hydrogen cooling edge at $\sim 10^4$ K is a good example).  It $\bf{also}$ generally helps for the jet to be significantly less dense than the cloud.  Finally, the jet velocity needs to be fast enough to trigger shocks in the cloud, yet not so fast that the cloud is disrupted before cooling can have much of an effect.

\section{Numerical Models}

Our numerical simulations are performed using the well-tested {\em Cosmos++} computational astrophysics code \citep{Anninos05}, specifically its Newtonian hydrodynamics solvers.  {\em Cosmos++} carries over many of the multi-physics capabilities found in its predecessor code {\em Cosmos} \citep{Anninos03}.  The Newtonian solvers have previously been utilized to study the bar mode instability in magnetized, rotating neutron stars \citep{Camarda09} and the galactic center G2 event \citep{Anninos12}.  The current work uses the High Resolution Shock Capturing (HRSC) scheme, which was described in its relativistic form in \citet{Fragile12}.  As there are few differences between the relativistic and Newtonian forms, we do not give a full presentation here, focusing instead on the packages that are most important to this paper: chemistry, cooling, and star formation. Note that, although magnetic fields can play an important role in shock-induced star formation \citep[cf.][]{Fragile05}, they are not considered in this work.

Most prior numerical studies of stimulated star formation from jet-generated shocks have been hampered by the resolution of the computational mesh \citep[e.g.][]{Fragile04}.  In the present work, we significantly improve on previous resolution limitations by employing the adaptive mesh refinement (AMR) capabilities of {\em Cosmos++}.  {\em Cosmos++} employs a {\em local} AMR scheme, in which refinement and de-refinement decisions are made on a cell-by-cell basis, using an oct-tree network to traverse the grid hierarchy \citep{Anninos05}.  Each level of refinement doubles the spatial resolution in each dimension within a given parent cell.  This style of local AMR scheme ensures that the refinement and de-refinement conform as closely as possible to the shape of the region of interest, in this case the shocks and unstable cooling fronts triggered inside the cloud by the jet. The AMR capabilities and the improvements in computing power have also allowed us to move from two-dimensional to more realistic three-dimensional simulations. Other improvements over our previous work include: simulating an object the size of M.O., instead of much smaller cloudlets; inputting a realistic jet, instead of a planar shock; inclusion of a star-formation prescription; and inclusion of dust grain chemistry.

In this work, we present an idealized case of a direct, axially symmetric collision between a jet and a pre-existing spherical cloud.  As such, it is a simple, and well controlled, test simulation, albeit with imperfect correspondence to M.O. In the case of M.O., it is thought that the collision was between a jet and a stellar bridge connecting the elliptical galaxy, NGC 541, with the interacting galaxies, NGC 545/547 (C06).  There is also evidence that the jet is slowly sweeping across M.O. (C06).  Despite these differences, the correspondence in parameters between our simulation and M.O. ensure that our results are applicable and the simulation can be used to better understand the dynamics of this particular object and of jet-induced star formation more generally.

\subsection{Simulation Setup}

Although we performed some two-dimensional simulations to test different code options and explore our parameter space, we focus on reporting the results of our 3D simulations.  The 3D simulations have a base resolution of $384\times128\times128$ zones to cover a domain that is approximately $30~\mathrm{kpc} \times 10~\mathrm{kpc} \times 10~\mathrm{kpc}$ with reflection boundaries applied in the $y$- and $z$-directions, so that we only simulate one quadrant of the full problem. The finest spatial resolution achieved is 19.5 pc per zone, reached by including 2 levels of refinement on top of the $384\times128\times128$ base mesh, equivalent to a uniform mesh of $1536\times512\times512$ zones.  The criterion used for refinement is that any zone with $n \ge 0.01$ cm$^{-3}$ is kept at the maximum refinement, while zones that fall below $n < 0.0005$  cm$^{-3}$ are allowed to de-refine, provided neighboring zones never differ by more than one level of refinement and no zone is allowed to drop below the base resolution. Zones are checked against the refinement and de-refinement criteria once every ten evolution steps in the numerical code.  We find that a minimum resolution close to our base value is required even in the background in order to get reasonable convergence in the star formation rate.

The cloud, which represents the parent object of M.O., is initialized with a radius of $R_\mathrm{cl} = 7.5$ kpc.  It is, therefore, somewhat smaller, in terms of projected area, than the real value of 275 kpc$^2$ (C06). The cloud is modeled as non-self-gravitating gas within a fixed 
dark-matter potential.  The omission of self-gravity is reasonable, given that the gravitational potential in an object like M.O. will be dominated by dark-matter \citep{Persic96}. The shape of the potential is given by a modified Hubble profile \citep{Binney87}
\begin{equation}
\phi(r<R_t) = \frac{G \tilde{M}}{R_c} \left\{ 1 - \frac{\mbox{ln}
[x + (1+x^2)^{1/2}]}{x} \right\} ~,
\end{equation}
where $x=r/R_c$, $R_c = 0.5 R_\mathrm{cl}$ is the core radius, and
\begin{equation}
\tilde{M} = M_d \left\{ \mbox{ln}[x_t +(1+x_t^2)^{1/2}] - 
x_t(1+x_t^2)^{-1/2} \right\}^{-1} ~,
\end{equation}
where $M_d = 10^{11} M_\odot$ is the dark-matter mass, $x_t=R_t/R_c$, and $R_t = 10 R_\mathrm{cl}$ is the tidal radius. 
The gas is initialized to be isothermal, with $T_\mathrm{cl} = 2 \times 10^5$ K, and in hydrostatic equilibrium within the potential, such that the density
\begin{equation}
\rho_\mathrm{cl} \propto e^{-\phi/c_s^2} ~,
\label{eqn:densitylaw}
\end{equation}
where $c_s$ is the isothermal sound speed. The gas within this potential is made clumpy by overlaying a random, log-normal distribution of the form $\tilde{n} e^{X\sigma}$, with standard deviation $\sigma = \sqrt{2 \ln (\bar{n}/\tilde{n})}=0.05$ and $\bar{n} = 0.5$ cm$^{-3}$, where $X$ is a randomly drawn variable with a mean of 0 and variance of 1. The normalization is such that the total gas mass within $R_\mathrm{cl}$ is $M_g = 1.4 \times 10^9 M_\odot$, which we show gives about the right mass of \ion{H}{1} ($4.9 \times 10^8 M_\odot$; C06), and giving an overall average density in the cloud of $\bar{\rho}_\mathrm{cl} = 5.3 \times 10^{-26}$ g cm$^{-3}$. The gas is initialized to have a mean molecular weight of $\mu = 1.3$, appropriate for neutral, solar metallicity gas, although this only affects the initialization, as $\mu$ is subsequently solved for self-consistently by the chemistry package. The cloud is immersed in a background gas with $n_\mathrm{b} = 10^{-4}$ cm$^{-3}$ and $T_\mathrm{b} = 5 \times 10^{7}$ K, such that the cloud and background are initially in approximate pressure equilibrium at the cloud surface. To avoid numerical problems caused by the background gas density dropping too far below its initial value, a density floor of $10^{-7}$ cm$^{-3}$ is set. Each simulation is run for a total of four sound-crossing-times, $R_\mathrm{cl}/c_{s,\mathrm{b}}$, of the cloud in the background gas, which corresponds to about 40 Myr total.

A jet is introduced into the simulation domain through one end. First, we estimate the power of the jet impacting M.O. from the claimed correlation with radio luminosity, $P_\mathrm{jet} \approx 7.2 \times 10^{36} (L_\mathrm{rad}/10^{30}~\mathrm{erg~s}^{-1})^{12/17} ~\mathrm{erg~s}^{-1}$ \citep{Kording08}, using the measured radio luminosity of NGC 541, $L_\mathrm{rad} \approx 10^{41}$ erg s$^{-1}$ \citep{vanBreugel85}. This gives a value for the kinetic power of $4 \times 10^{44}$ erg s$^{-1}$. This estimate is roughly consistent with independent estimates of the jet power from the other radio source in Abell 194, 3C40B, based on X-ray cavity energetics \citep{Bogdan11}. Since both sources have similar radio luminosities, it seems reasonable that their jet powers would be similar, too. In order to translate this jet power into simulation variables, we are also guided by dynamical models for jet deceleration in intergalactic environments \citep[e.g.][]{Laing02}. We settle on the following jet parameters: a density of $n_\mathrm{jet} = 5 \times 10^{-6}$ cm$^{-3}$, a diameter of $D_\mathrm{jet} = 5$ kpc, and a velocity of 10\% of the speed of light. For a cylindrical jet of cross-sectional area $A_\mathrm{jet}$, this yields a kinetic power, $P_\mathrm{jet} = \rho_\mathrm{jet} A_\mathrm{jet} v_\mathrm{jet}^3 = 6 \times 10^{43}$ erg s$^{-1}$, somewhat below the power estimate above, consistent with the jet having dissipated some fraction of its power prior to reaching M.O.

\subsection{Chemistry Models}

In our simulations, we follow the abundances of 9 atomic and molecular species: \ion{H}{1}, \ion{H}{2}, \ion{He}{1}, \ion{He}{2}, \ion{He}{3}, $e^-$, H$^-$, H$_2$, and H$_2^+$. The evolution of each species is governed by an equation of the form
\begin{equation}
\frac{\partial \rho^{[m]}}{\partial t} +\nabla \cdot (\rho^{[m]} \mathbf{v})
        = {\sum_{i=1}^{N_{s}}}{\sum_{j=1}^{N_{s}}} 
                            {{k_{ij}(T)}{\rho^{[i]}}{\rho^{[j]}}}
            + \sum_{i=1}^{N_{s}} {I_i (\nu) \rho^{[i]}} e^{-\tau_i}.
\label{eqn:dens_m}
\end{equation}
These rate equations are solved using a stable, semi-implicit, backward difference scheme that we developed in \citet{Anninos97}, which has since become a standard method adopted by the general community \citep[e.g.][]{Smith16} due to its combination of robustness, efficiency, and accuracy.  A total of 27 gas-phase chemical reactions are included in the full network,  including 19 collisional ($k_{ij}$) and 8 photoionization/photodissociation ($I_i$) processes.  The exact reaction chains are spelled out in \citet{Anninos03}.  The photoionization field is set to $10^{-21}$ s$^{-1}$, appropriate for cosmic UV background radiation at low redshift \citep{Bechtold87}, while the photodissociation rate is $5.0 \times 10^{-11}$ s$^{-1}$, appropriate for the local interstellar medium \citep{Spaans97}. For photoionization, the products of the external field and the respective interaction cross sections are each integrated over frequency to derive effective photoionization rates for H, He, and molecules. To account for self-shielding within the cloud, we approximate the optical depth as $\tau_i = \sigma_i n_i \Delta l$, where $\sigma_{\mathrm{H}_\mathrm{I}} = 6.3 \times 10^{-18}$ cm$^2$ \citep{Osterbrock89}, $\sigma_{\mathrm{H}_2} = 5.2 \times 10^{-18}$ cm$^2$ \citep{Hollenbach71}, and $\Delta l$ is the length of a typical zone.

We account for the effect of dust grains by adding an extra reaction to the network consisting of collisional interactions between grains and hydrogen atoms to enhance the production of molecules. Dust grains can also absorb and emit radiation, effectively acting as an additional cooling or heating mechanism. Both collisional and cooling rates are sensitive functions of the gas and grain temperatures, and of the grain size.
We adopt the grain reaction and cooling models of \cite{Hollenbach79} and \cite{Omukai00} for this work, assuming a characteristic grain size and temperature of 100 {\AA} and 10 K, respectively.

\subsection{Cooling Models}

The energy equation 
\begin{equation}
\frac{\partial E}{\partial t} +\nabla \cdot [(E+P) \mathbf{v}]
        = -\Lambda(T,n^{[m]})~,
\label{eqn:dual_en}
\end{equation}
where $E = e + \rho v^2/2$ is the total energy density, including its internal and kinetic contributions, accounts for the cooling and heating of the gas via a total of eight different mechanisms: collisional-excitation, collisional-ionization,
recombination, bremsstrahlung, metal-line cooling (dominantly carbon, oxygen, neon, and iron), molecular-hydrogen cooling, dust cooling, and photoionization heating. The cumulative cooling function is
\begin{equation}
\Lambda(T,n^{[m]}) =
    \sum_{i=1}^{N_{s}} \sum_{j=1}^{N_{s}}  \dot{e}_{ij}(T) n^{[i]} n^{[j]}
    - \sum_{i=1}^{N_{s}} J_i n^{[i]}
    + \dot{e}_M(T) n^2 ~,
\label{eqn:coolingchem}
\end{equation}
where $\dot{e}_{ij}(T)$ are the cooling rates from 2-body interactions between species $i$ and $j$, and $J_i$ represents the frequency-integrated photoionization and photodissociation heating rates. $\dot{e}_M$ is the temperature-dependent cooling rate for metals (assuming solar metallicity), taken from \citet{Dalgarno72}. A cooling floor is set at $T_\mathrm{floor} = 10$ K, below which only adiabatic cooling is possible.

\subsection{Star Formation Model}

In this study, we are particularly interested in tracking the formation of stars within the cloud.  We follow the approach of \citet{Rasera06} in defining a density threshold, $n_\star$, above which star formation is triggered at a rate given by
\begin{equation} 	
\dot{\rho}_\star = \frac{\epsilon \rho}{t_\mathrm{ff}} 
\label{eqn:sfr} 
\end{equation}
where $t_\mathrm{ff} = \sqrt{3 \pi/32 G \rho}$ is the local free-fall time and $\epsilon$ controls the star formation efficiency. We tested values of $n_\star$ between 0.5 and 4 cm$^{-3}$ and $\epsilon$ between 0.02 and 0.1, settling in our two highest resolution simulations on $n_\star = 1$ cm$^{-3}$ and $\epsilon = 0.02$. This star formation appears as a sink term in the continuity equation
\begin{equation}
\frac{\partial \rho}{\partial t} +\nabla \cdot (\rho \mathbf{v}) = -\dot{\rho}_\star ~.
\label{eqn:dens}
\end{equation}
By tracking how much mass is lost to star formation during each compute cycle, we are able to continuously track the star formation rate throughout the simulation. 

Additionally, we use tracer particles to track the most massive ($M_\star \ge 0.02 M_\odot$) ``stars'' created in this way.  Each star particle is given an initial velocity equal to the velocity of the gas in the zone in which the star is created. The particles are then fed into a post-processing routine, which integrates their motion through the dark-matter potential. This is not entirely realistic as some momentum may get redistributed during the free-fall process that leads to star formation, but we at least capture the dominant force that would act on the stars once they form.

We are justified in our neglect of self-gravity in following the collapse of the cloud, because the initial Jeans radius
\begin{equation}
R_J = \left(\frac{15 kT}{4\pi G\mu m_H \rho}\right)^{1/2}
\end{equation}
of the cloud is 21 kpc, more than twice the size of the cloud. Even at the densities and temperatures typical of star formation in our simulations ($\rho_\star \approx 2 \times 10^{-24}$ g cm$^{-3}$ and  $T_\star \approx 1000$ K), the Jeans length is still 240 pc, which is considerably less than the original radius of the cloud, yet well above the resolution limits of our simulations. More importantly, it is comparable to the size scales of the regions that exceed the star formation criteria, meaning the clumps are just becoming Jeans unstable whenever our star formation model kicks in and starts converting gas in these regions to stars. In other words, just when self-gravity would be taking over is when our star formation model kicks in.

\section{Results}

\begin{figure}
\includegraphics[width=0.75\columnwidth]{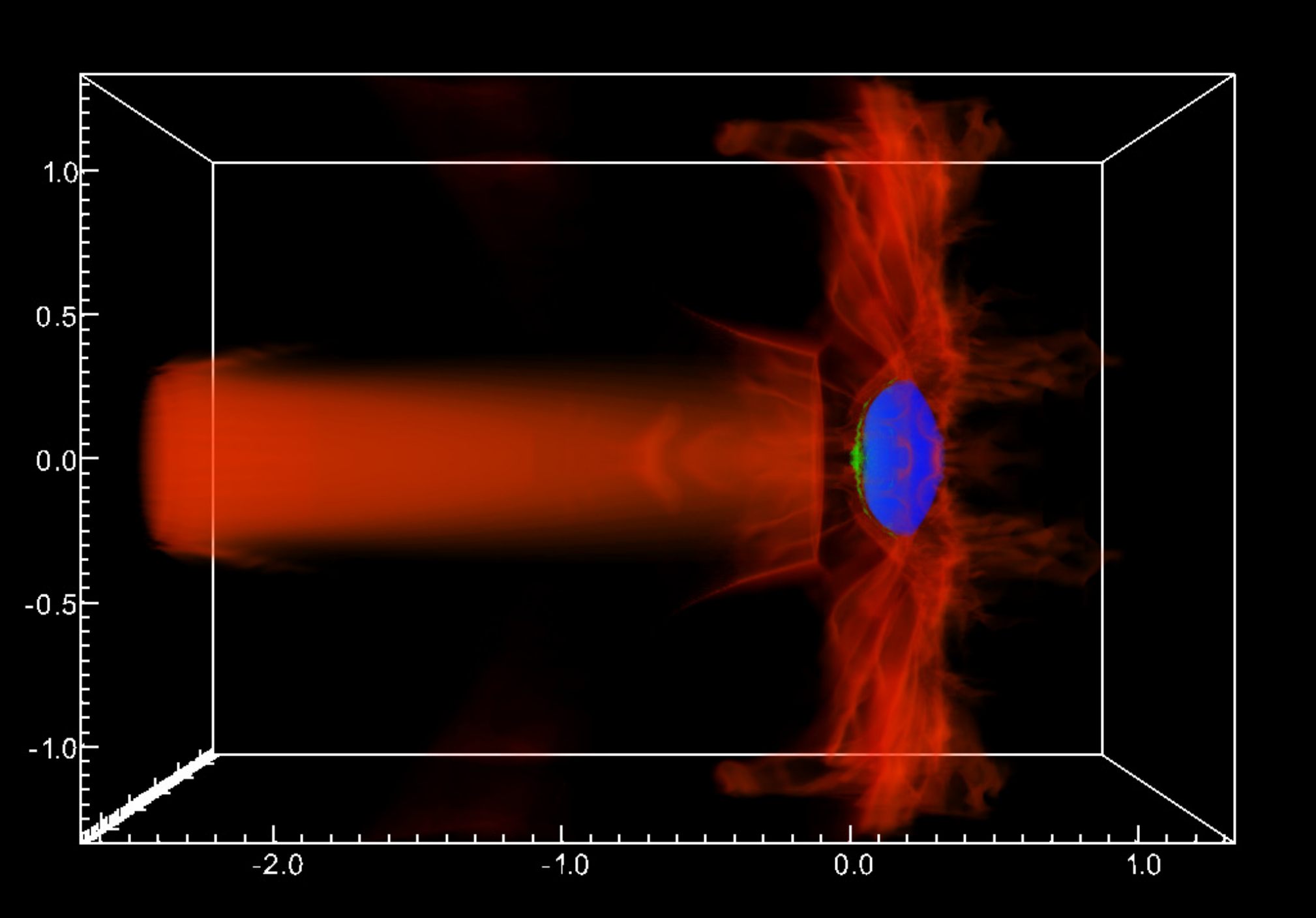} 
\includegraphics[width=0.75\columnwidth]{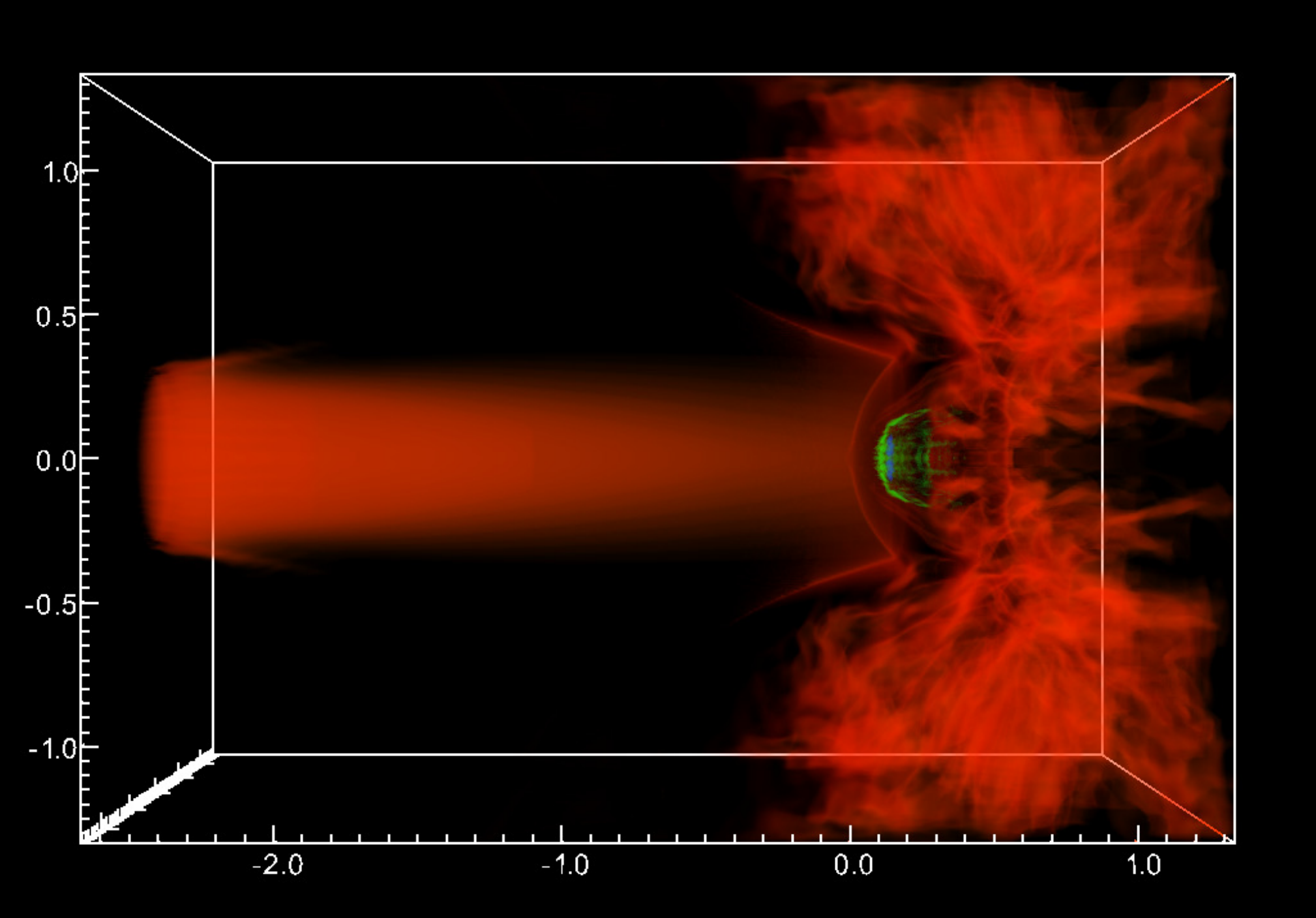} 
\caption{Volume visualization from an intermediate ($t=30$ Myr) and the final ($t=40$ Myr) time dumps of our highest resolution simulation (384x128x128\_3level).  Red represents hot, $T > 2\times 10^8$ K, jet material; blue represents regions of the cloud with a neutral hydrogen density, $n_\mathrm{HI} \gtrsim 0.1$ cm$^{-3}$; and green represents regions of the cloud with a molecular hydrogen (our tracer for cold gas) density, $n_\mathrm{H2} \gtrsim 10^{-5}$ cm$^{-3}$.  Data have been reflected across the $y=0$ and $z=0$ planes to create this image.  Axes are marked in units of cloud radii, $R_\mathrm{cl}$.
\label{fig:volume}}
\end{figure}

Table \ref{tab:params} summarizes the 3 simulations presented in this work.  The naming convention refers to the base resolution and how many total grid resolution levels there are.  Figure \ref{fig:volume} presents volume visualizations from an intermediate- and the end-time of our highest resolution 3D simulation (384x128x128\_3level).  A number of general features are apparent.  First, the cloud is dense enough and has enough inertia to dramatically slow the propagation of the jet (represented by its temperature in red).  At the speed the jet is traveling, if not for the cloud (and, to a lesser extent, the background gas) impeding its progress, it should have traversed 41 box lengths (164 cloud radii) over the duration of the simulation.  Instead, most of the jet material is deflected to the sides of the cloud, though a significant fraction of its energy is deposited within the cloud gas.  The blue material represents neutral hydrogen (\ion{H}{1}) and shows that the core of the original cloud remains relatively intact until late times. At the head of the jet, where it is interacting with the cloud, a thin layer of very dense, cold gas has formed (represented by the green, H$_2$, gas). This is where star formation is expected to occur. This figure shows qualitative similarities with Fig. 1 of \citet[][hereafter L17]{Lacy17}, which reports ALMA observations of M.O. 

\begin{deluxetable}{cccc}
\tablecaption{Jet-Cloud Models and Parameters \label{tab:params}}
\tablecolumns{4}
\tablehead{
 &  & \colhead{$n_\star$} & \\
\colhead{Name\tablenotemark{a}} & \colhead{$N_l$\tablenotemark{b}} & \colhead{(cm$^{-3}$)} & \colhead{$\epsilon$} }
\startdata
384x128x128 & 1 & 1 & 0.1  \\
384x128x128\_2level & 2 & 1 & 0.02  \\
384x128x128\_3level & 3 & 1 & 0.02  \\
\enddata
\tablenotetext{a}{Each simulation has a base resolution of $384\times 128\times128$.}
\tablenotetext{b}{$N_l$ is the total number of grid refinement levels.}
\end{deluxetable}

\subsection{Shock Propagation}
\label{sec:shock}

In Figure \ref{fig:volume}, we can clearly see the termination shock at the end of the collimated jet. Ahead of it is a compression shock being driven into the cloud (not seen in Figure \ref{fig:volume}, but lying just ahead of the cold, dense H$_2$ gas seen in green). We can compare its position with the estimated shock velocity in the cloud. If we take our jet velocity, $v_\mathrm{jet} = 3 \times 10^4$ km s$^{-1}$, as the speed of the post-shocked material inside the jet, then we can use the usual jump conditions, 
\begin{equation}
v_\mathrm{ps} = \left(1 - \frac{\Gamma-1}{\Gamma+1}\right) v_\mathrm{sh} ~,
\end{equation}
where $v_\mathrm{ps}$ and $v_\mathrm{sh}$ are measured in the rest frame of the pre-shock gas, to estimate the speed of the jet shock, $v_\mathrm{sh,jet} \approx 4 v_\mathrm{jet}/3 = 4 \times 10^4$ km s$^{-1}$. If the shock in the jet is strong, then the post-shock pressure is approximately $\rho_\mathrm{jet} v_\mathrm{sh,jet}^2$. If we assume the shocks are also strong inside the background and that the post-shock jet and background gas reach pressure equilibrium, then we can estimate the speed of the shock in the background 
\begin{equation}
v_\mathrm{sh,b} \simeq \left(\frac{\rho_\mathrm{jet}}{\rho_\mathrm{b}}\right)^{1/2} v_\mathrm{sh,jet} = 8.9 \times 10^3 ~\mathrm{km\,s}^{-1}.
\end{equation}
If we likewise assume the shocks are strong in the cloud and take a characteristic cloud density of $\rho_\mathrm{cl} = 2.2 \times 10^{-25}$ g cm$^{-3}$, we get a shock speed in the cloud of
\begin{equation}
v_\mathrm{sh,cl} \simeq \left(\frac{\rho_\mathrm{b}}{\rho_\mathrm{cl}}\right)^{1/2} v_\mathrm{sh,b} = \frac{v_\mathrm{sh,b}}{\chi^{1/2}} = 280 ~\mathrm{km\,s}^{-1},
\end{equation}
where $\chi = \rho_\mathrm{cl}/\rho_\mathrm{b}$ is the ratio of cloud to background density. According to this, the shock should traverse the cloud core in about 26 Myr, which looks to be roughly consistent with the shock progression seen in Figure \ref{fig:volume}.

\subsection{Cooling Front}
\label{sec:cooling_front}

\begin{figure}
\includegraphics[height=0.5\columnwidth]{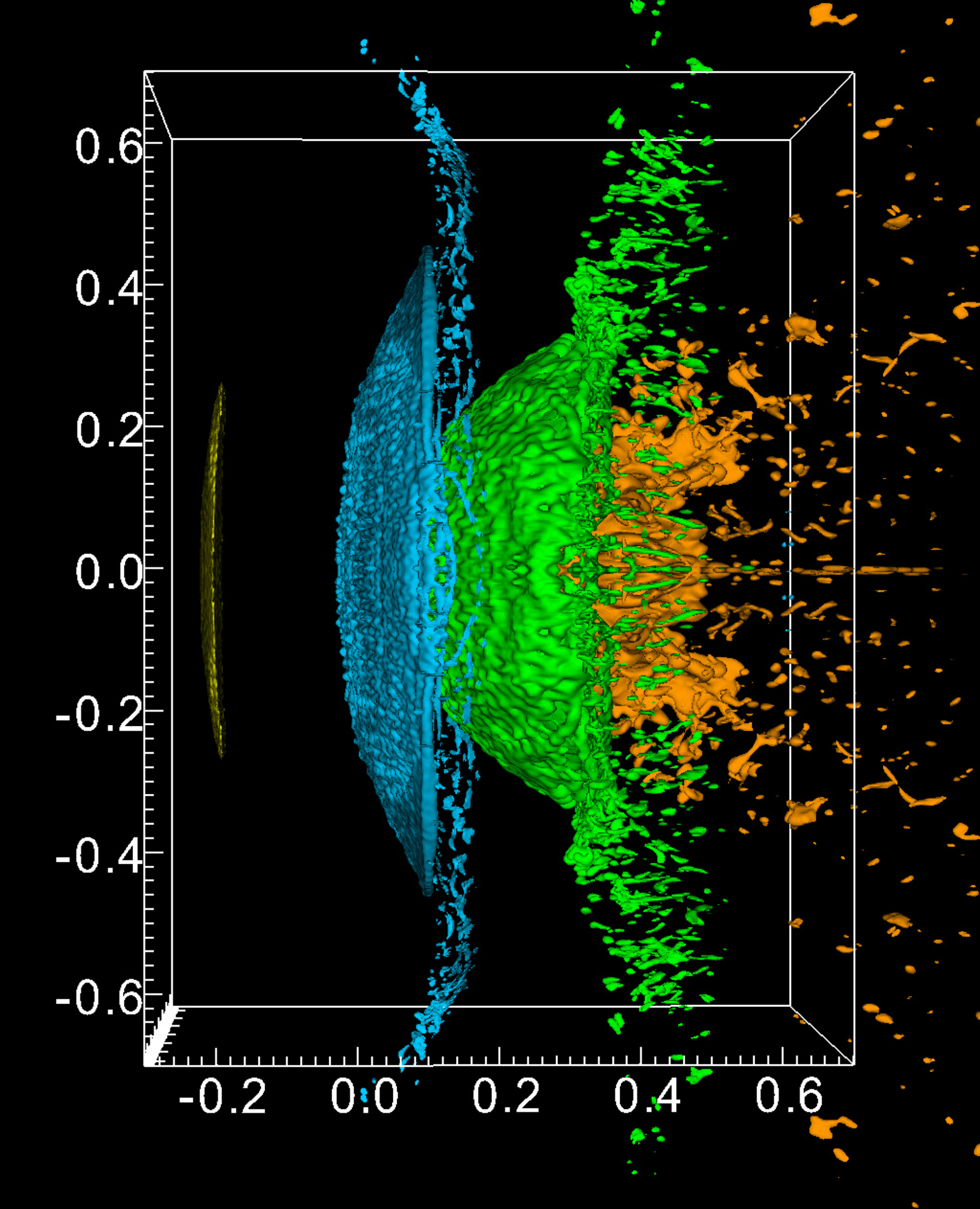} 
\caption{Isosurface plot showing regions of the simulation domain where the density exceeds the star formation limit ($n > n_\star$) at $t = 10$ (yellow), 20 (blue), 30 (green), and 40 Myr (orange) from our highest resolution simulation (384x128x128\_3level). Data have again been reflected across the $y=0$ and $z=0$ planes. Axes are marked in units of cloud radii, $R_\mathrm{cl}$. 
\label{fig:formation}}
\end{figure}

As mentioned in Section \ref{sec:jetcloud}, for the jet feedback to be positive, it is critical for the cooling timescale to be shorter than the disruption timescale of the cloud, taken to be the shock-crossing time. Following \citet{Fragile04}, we estimate the cooling time to be
\begin{equation}
t_\mathrm{cool} = (7.0 \times 10^{-35}~\mathrm{g~cm}^{-6}~\mathrm{s}^4) \frac{v^3_\mathrm{sh,cl}}{\rho_\mathrm{cl}} \simeq 2.2 \times 10^{5}~\mathrm{yr} ~.
\end{equation}
This is {\it much} shorter than any other relevant timescale in the problem. Figure \ref{fig:formation} shows how this cooling front and the associated star formation progresses over the course of the simulation.

\subsection{Star Particles}
\label{sec:star_particles}

\begin{figure}
\includegraphics[width=0.5\columnwidth]{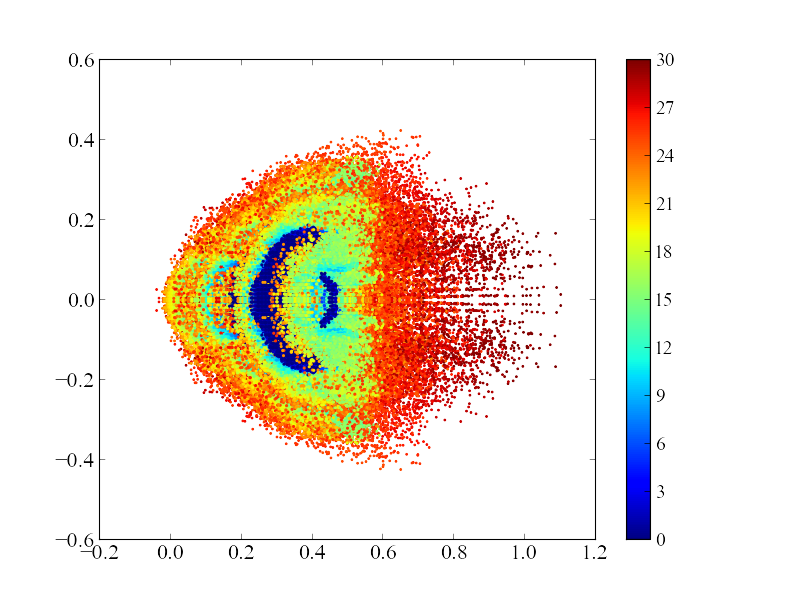} 
\includegraphics[width=0.5\columnwidth]{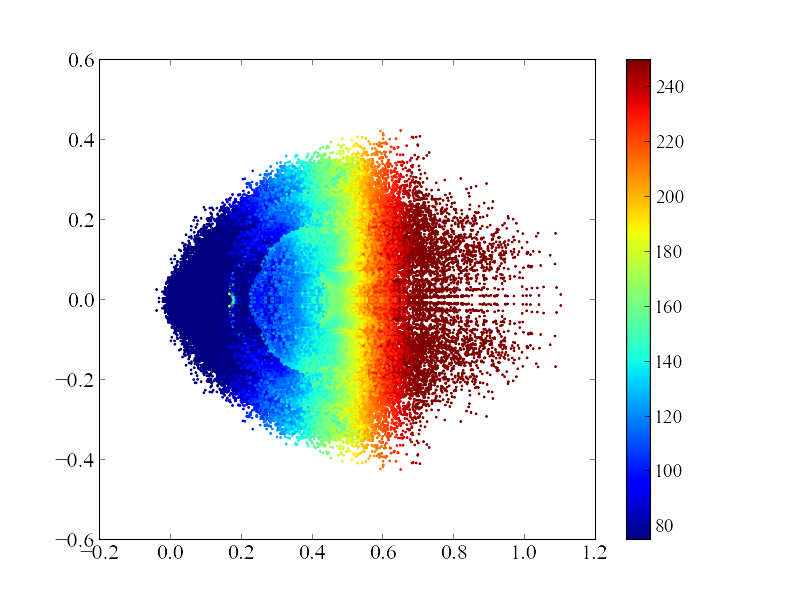} 
\caption{Plot of spatial distribution (projected into the $x$-$y$ plane) of star tracer particles from the final time dump ($t=40$ Myr) of our lowest resolution 384x128x128 simulation.  In the left panel, the color of each particle indicates its age (in Myr), while in the right panel, the color indicates velocity magnitude (in km s$^{-1}$). 
\label{fig:stars}}
\end{figure}

As a reminder, whenever more than $0.02 M_\odot$ of gas is converted into stars within a given zone within a single cycle, then a star particle is created to track the properties, such as position, age, and velocity, of that ``star.''  Following this prescription, we created over $2.3 \times 10^7$ star particles in our lowest resolution simulation (384x128x128).  The star particles span an age range from 0 -- 32 Myr. However, this ``age'' does not correspond directly to the age of a star, as we do not account for the freefall and pre-main-sequence lifetimes of each. At best, the star particle ages give an estimate of the range of ages that may be expected and, as shown in Figure \ref{fig:stars}, some feeling for the spatial distribution of ``young'' and ``old'' stars.

An interesting point about the stellar ages in our simulations is that they show a negative curvature along the direction of jet propagation, that is, the youngest (currently forming) stars are found between two populations of slightly older stars, one in the upstream direction and one downstream. The upstream population are stars that formed recently, only slightly before the current star formation.  The downstream population are some of the first stars to form from the jet interaction, but they are now actually located ahead of the current star formation front because they received a velocity kick larger than the current shock speed.  Since the star particles are not coupled to the gas, they can actually pass ahead of the shock as it slows down, giving an apparently older population ahead of the current star formation front.  

\begin{figure}
\includegraphics[width=0.5\columnwidth]{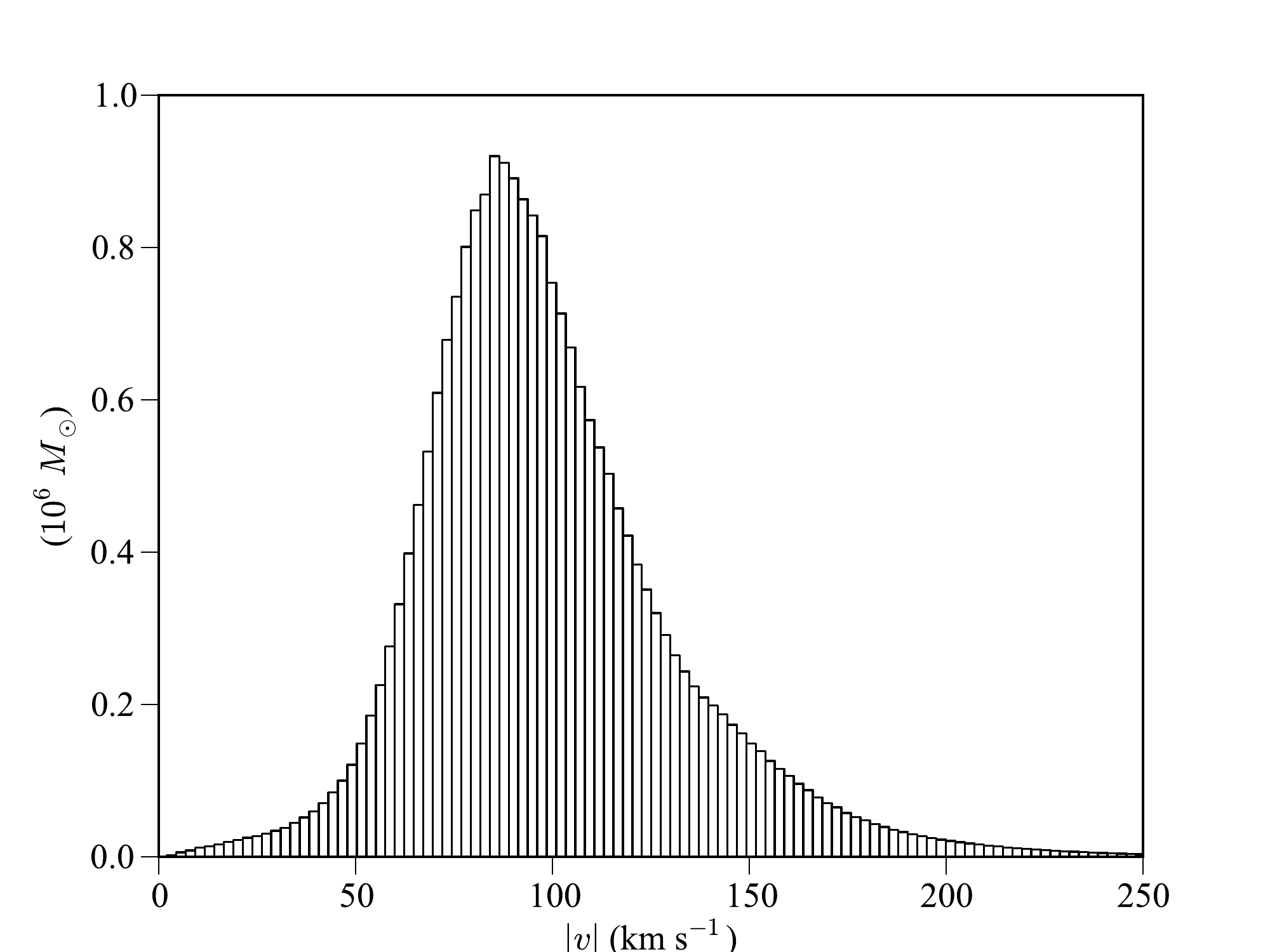} 
\caption{Histogram of the velocity magnitudes of all the star particles formed in our 384x128x128 simulation. The peak of the distribution is roughly consistent with observations of M.O.
\label{fig:histogram}}
\end{figure}

Figure \ref{fig:stars} (right panel) shows the spatial distributions of the star particles, colored by the magnitude of their velocities.  As mentioned above, the fastest moving star particles are on the downstream edges of the distribution. By using our results from Sec. \ref{sec:shock} and inverting the shock jump condition, we predict a post-shock velocity in the cloud of $v_\mathrm{cl,ps} = 3 v_\mathrm{sh,cl}/4 \simeq 210$ km s$^{-1}$. Since our star particles are assigned their velocity based upon the velocity of the gas from which they form, we expect the simulated star particle velocities to be similar.  Figure \ref{fig:histogram} shows a histogram of the velocity distribution of all of the star particles. The distribution peaks at around 85 km s$^{-1}$, which is fairly consistent with our crude predictions. The spread in our velocity distribution is also fairly consistent with the $\sim 40$ km s$^{-1}$ of velocity shear observed in M.O. (L17).

\subsection{Star Formation Rate}
\label{sec:sfr}

\begin{figure}
\includegraphics[width=0.5\columnwidth]{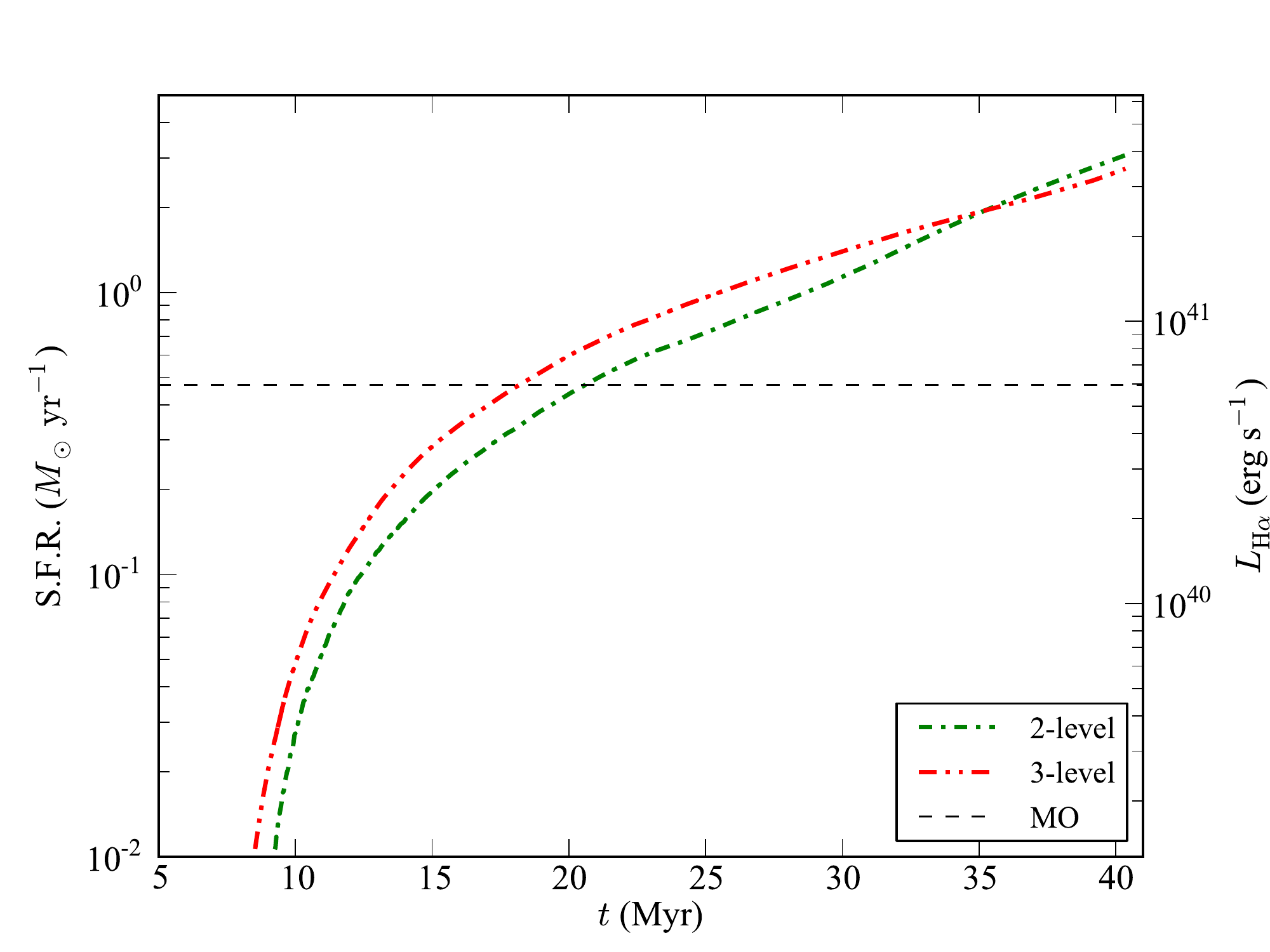} 
\caption{Star formation rate history for the two highest resolution simulations. After about 20 Myr, the S.F.R. in both simulations exceeds the current observed rate in M.O. of $\approx 0.5 M_\odot$ yr$^{-1}$ (grey, dashed line, corresponding to an H$_\alpha$ luminosity of $7 \times 10^{40}$ erg s$^{-1}$).
\label{fig:sfr}}
\end{figure}

Figure \ref{fig:sfr} shows the measured star formation rate (S.F.R.) as a function of time for two different 3D simulations done at different effective resolutions, from 39.1 pc per zone (384x128x128\_2level) to 19.5 pc per zone (384x128x128\_3level). Up to a point, we expect the simulated S.F.R. to be sensitive to resolution. At higher resolutions, more gas is going to be able to reach the density threshold, $n_\star$, of our star formation model. However, a limit should be reached where the high density filaments are well enough resolved that their size and peak density are no longer functions of resolution. It appears we may have reached this point in our highest resolution simulations, as they track each other closely. Equally important, we achieved an S.F.R., or correspondingly an H$_\alpha$ luminosity, since \citep{Kennicutt98}
\begin{equation}
\left(\frac{\mathrm{SFR}}{M_\odot~\mathrm{yr}^{-1}} \right) = 7.9 \times 10^{-42} \left(\frac{L_{H_\alpha}}{\mathrm{erg~s}^{-1}}\right) ~,
\end{equation}
consistent with the observed value in M.O. \citep[$\mathrm{S.F.R.} = 0.47 M_\odot$ yr$^{-1}$ or $L_{H_\alpha} = 5.9 \times 10^{40}$ erg s$^{-1}$;][]{Salome15}.  In fact, we overshoot the observed S.F.R. after about 20 Myr. However, we are neglecting negative feedback effects, such as supernovae from the first generation of stars, which would damp this rate.

\subsection{Other Comparisons with Observations}

\begin{figure}
\includegraphics[width=0.5\columnwidth]{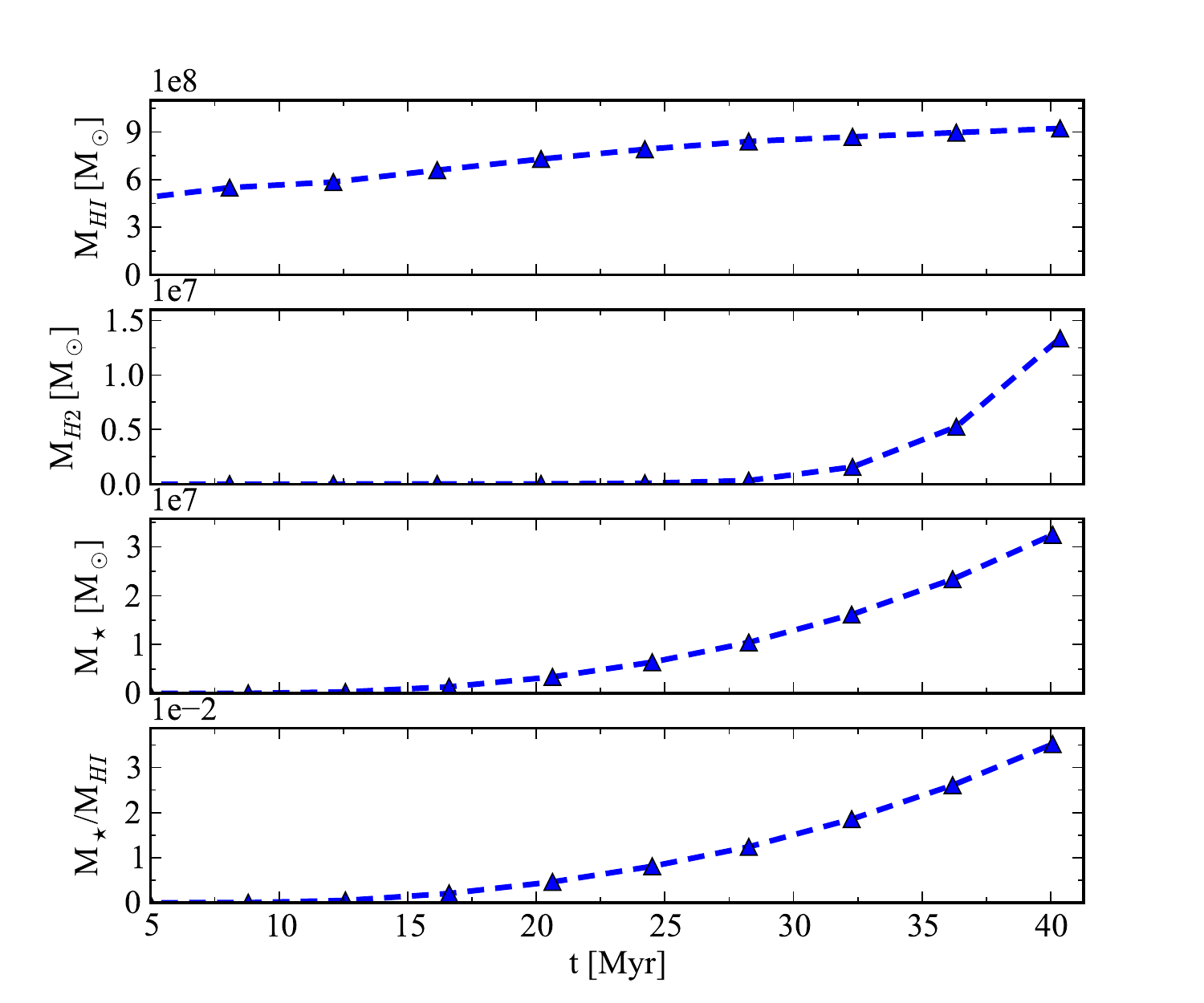} 
\caption{Plots of $M_\mathrm{HI}$, $M_{\mathrm{H}_2}$, $M_\star$, and $M_\star/M_\mathrm{HI}$ over time for our highest resolution 3D simulation (384x128x128\_3level). At the final time of $t = 40$ Myr, all measures are within a factor of two of their observed values.
\label{fig:history}}
\end{figure}

Figure \ref{fig:history} tracks the total mass of \ion{H}{1}, H$_2$, and stars over the course of our highest resolution simulation. All of these measures are within about a factor of two of their observed values. The simulation slightly overproduces \ion{H}{1} [$9.2 \times 10^8 M_\odot$ in the simulation vs. $4.9 \times 10^8 M_\odot$ for M.O. (C06)], slightly underproduces H$_2$ [$1.3 \times 10^7 M_\odot$ vs. $(3.0-18) \times 10^7 M_\odot$ in M.O. (L17)], and slightly overproduces stars [$3.3 \times 10^7 M_\odot$ vs. $1.9 \times 10^7 M_\odot$ in M.O. (C06)].  We find an electron number density in our cloud of $n_e \sim 0.2$ cm$^{-3}$, also somewhat lower than the range of 1-10 cm$^{-3}$ obtained for M.O. (C06). Figure \ref{fig:history} also tracks the ``star formation efficiency,'' $M_\star/M_\mathrm{HI}$, over time.  By this measure, our simulation achieves a peak star formation efficiency of 3.5\%, very close to the value of 4\% measured in M.O. (C06). 

\section{Discussion \& Conclusion} 

In this work, we used 3D multi-physics hydrodynamic simulations to model the evolution of a radio jet impacting a single, dwarf-galaxy-scale cloud in a direct, axially-symmetric collision. Our intention is to use these simulations to better understand observations of M.O., the peculiar starburst galaxy located at the termination point of the radio jet from NGC 541. 

Our first, main conclusion is that jet-induced star formation (i.e. positive feedback) is possible under this scenario. Figure \ref{fig:sfr} shows a dramatic increase in star formation attributable to the jet interaction, and the star-formation rate matches M.O. well.  Importantly, this star formation is occurring upstream of the bulk of the \ion{H}{1} (compare Figures \ref{fig:volume} and \ref{fig:formation}), also consistent with observations (C06). 

Other quantitative measures from the simulations also show substantial agreement with the observations. For example, the total masses of \ion{H}{1}, H$_2$, and stars (Figure \ref{fig:history}) are all within a factor of two of their M.O. values. Additionally, the velocity histogram of our star particles peaks at $| v | \approx 85$ km s$^{-1}$ (Figure \ref{fig:histogram}), which is roughly consistent with observations (L17). Given the relative simplicity of our setup (uniform jet hitting spherical cloud head on), it is remarkable how well our results match quantitatively across such a wide range of diagnostics.

The spatial distribution of the star particle velocities is interesting. The fastest moving star particles are found furthest downstream (Figure \ref{fig:stars}). These are also some of the first star particles to form (i.e. they are the oldest), which is consistent with a slowing of the propagation speed of the star formation front. We plan to reexamine the observations of M.O. to see if a similar distribution is present.

One effect that is not treated in these simulations is negative feedback from the star formation process itself. Once the first generation of massive stars form, there is only a limited amount of time (of the order a few Myr) for star formation to continue before heating from these stars would effectively shut it off \citep{Dong03}. This might explain why the S.F.R. in M.O. appears to be lower now than it was in the past -- negative feedback may already be kicking in.

As future observations continue to constrain the star formation history of M.O., we plan to continue to refine our simulations.  Future modifications to our setup may include: simulating non-axially-symmetric interactions between the jet and cloud; having the jet sweep across the cloud; or adding more feedback mechanisms, such as heating from young stars and supernovae. We could also include self gravity, which would only enhance and accelerate the star formation.

\acknowledgements
This work used the Extreme Science and Engineering Discovery Environment (XSEDE), which is supported by National Science Foundation grant number ACI-1053575.   PCF acknowledges support from National Science Foundation grants AST-1211230 and AST-1616185.  JWLW acknowledges support from NRAO Student Observing Support grant SOSPA3-020. Work by PA was performed in part under the auspices of the U.S. Department of Energy by Lawrence Livermore National Laboratory under Contract DE-AC52-07NA27344.

\software{Cosmos++ \citep{Anninos05}}

\end{document}